\documentclass[aps,superscriptaddress]{revtex4}
\usepackage{epsfig}
\usepackage{amsbsy}
\usepackage{amsmath}
\usepackage{graphicx}
\usepackage{latexsym}
\usepackage{amsfonts}
\usepackage{amsthm}
\usepackage{bm}

\newcommand\ket[1]{\left\vert#1\right\rangle}

\begin{document} 

\title{Non-Gaussian states for continuous variable
       quantum computation via Gaussian maps}
\author{Shohini Ghose}
\affiliation{Wilfrid Laurier University, Waterloo Ontario N2L 3C5, Canada}
\author{Barry C. Sanders}
\affiliation{Institute for Quantum Information Science, University of Calgary, Alberta T2N 1N4, Canada}

\date{\today}

\begin{abstract}
We investigate non-Gaussian states of light as 
ancillary inputs for generating nonlinear transformations required for 
universal quantum computing with continuous variables. 
We consider a recent proposal for preparing 
a cubic phase state, find the exact form of the prepared state and perform a detailed comparison to the ideal cubic phase state. We thereby identify the main challenges to preparing an ideal cubic 
phase state and describe the gates implemented with the non-ideal 
prepared state. We also find the general set of gates that can be 
implemented with ancilla  Fock states, together with Gaussian input states, linear optics and squeezing transformations, and homodyne detection with feed forward.  Such circuits can be used to approximate a certain class of  Hamiltonian evolutions. Furthermore, we analyze the question of 
efficient classical simulation of these gates. These results extend the 
existing theorems about efficient classical simulation for continuous variable quantum information 
processing.
\end{abstract}
\maketitle

\section{Introduction}
Universal quantum computation can be performed by encoding qubits in discrete states of quantum systems and performing a set of universal quantum logic gates or unitary operations on the qubits. For the case of light, the electromagnetic field amplitudes are described by operators that have a continuous rather than discrete spectrum. Quantum information processing with continuous variables (CV) is thus appealing because of the possibility of optical implementation using simple linear optics elements and coherent states of light. 
Furthermore, a CV quantum computer may be more suitable for simulating CV quantum systems. Whereas much progress has been made in areas of communication, measurement and computation using optical systems~\cite{CV,CV2}, universal CV quantum computation with light remains a challenge due to the weakness of optical nonlinearities. This problem can be circumvented by off-line preparation of nonclassical states of light that are fed into the optical circuit at various points to generate the required strong nonlinearities. This idea was
explored by Knill, Laflamme and Milburn~\cite{KLM01} and subsequently by others~\cite{YR03,N04, BR05} for quantum computing with qubits,
and by Gottesman, Kitaev, and Preskill (GKP) for qudits encoded in CV systems~\cite{GKP01}.
In this paper we present an analysis of non-Gaussian ancilla states of light in optical circuits as an off-line resource for generating universal gates for continuous variable quantum information processing.

For universal CV quantum information processing, the aim is to apply a finite set of gates that, by repeated applications, will generate unitary operations corresponding to a general class of Hamiltonian evolution. Specifically, we are interested in Hamiltonians that are arbitrary polynomial functions of the continuous variables~\cite{LB99}. In optical implementations, combinations of linear optics and squeezing elements can generate transformations that are limited to quadratic functions of the  field amplitude operators $\hat x$ and $\hat p$. Furthermore, any circuit consisting of input Gaussian states and these transformations, along with homodyne detection and feed forward can be efficiently simulated on a classical computer~\cite{BSBN02}. This 
efficient simulation is due to the fact that the process can be described as a Gaussian-preserving map and can be completely characterized by the evolution of the means and covariances of the initial Gaussian state. This requires performing simple operations on a finite set of variables, that scales polynomially with the number of inputs~\cite{BSBN02,BS02}.

Nonlinear transformations,
namely those outside the set of linear and quadratic operations,
are therefore required for universal CV quantum computing - i.e. for implementing 
unitaries corresponding to Hamiltonians that are polynomial functions of the continuous variables. One possibility is to use photon counting measurements to generate nonlinearities~\cite{SB02}. 
Photon counting plays a key role in schemes for linear optics quantum computation using qubits~\cite{KLM01} and for computation with CV-encoded qudits~\cite{GKP01}. The measurements can be used to generate non-Gaussian states through an `off-line' preparation step that is independent of the particular circuit being implemented.  
GKP then showed that these non-Gaussian states can be employed as ancillae at different stages of `on-line' quantum information processing to replace the
requirement for nonlinear gates. This partition of the quantum computation into off-line and on-line elements obviates the requirement for real-time photon-counting measurements that must adapt to the vagaries of
each specific circuit.

Here we investigate and analyze off-line preparation of non-Gaussian ancilla states 
and the employment of such ancilla states in on-line circuits. We specifically consider cubic phase states and Fock states. 
Our interest focuses on these off-line prepared states as resources for replacing
the nonlinearity  in order to effect universal quantum information processing. We first perform a detailed analysis of the scheme proposed by GKP for off-line preparation of a cubic phase state. We find the exact form of the state generated by the GKP scheme and show that it is a finite superposition of displaced number states. 
We use the fidelity as a figure of merit to assess the efficacy of preparing an ideal cubic phase state with this scheme, and present detailed studies of the Wigner function for experimental parameters that take into account finite squeezing and finite detector efficiency. Our analysis unfortunately indicates that current levels of squeezing are insufficient to prepare close approximations to the cubic phase state, although this limited squeezing does not necessarily  rule out this approximate
cubic phase state from providing an off-line resource for universal quantum information
processing with Gaussian operations in the on-line circuit. 
Our results are of importance to current experimental research with squeezed light and in particular, it provides guidelines to those actively working on experimentally preparing the cubic phase state~\cite{QCMC}. This work is also of relevance to recent studies of quantum computing with continuous variable cluster states~\cite{CVcluster}.

We also consider an alternative non-Gaussian state as an off-line
resource for universal quantum information processing with Gaussian elements in the
on-line circuit, namely the ubiquitous Fock state.
Whereas the cubic phase state is special in that it both delivers an off-line
resource for universal quantum information processing with Gaussian circuit elements
and also makes the incorporation of the cubic phase state into the circuit simple and 
elegant, the Fock state is also non-Gaussian and perhaps the most feasible to prepare,
at least for single-photon states~\cite{GSV05}.
Since the Fock state is probably the most amenable non-Gaussian
state to feed into a Gaussian circuit as an off-line resource, it is important to assess its status as a useful
off-line resource for CV computation.

In this analysis of Fock state resources,
we consider a Gaussian-preserving circuit receiving ancillary photons.
We determine the general form of non-Gaussian output states that can be obtained from a Gaussian-preserving circuit with an $n$-channel input state along with some ancillary channels, each occupied by a number
state. Our studies show that ancilla Fock states appear to provide the requisite nonlinearity for universal CV computation, but circuit design for implementing nonlinear gates is not obvious with Fock state inputs.
We also include a brief discussion of efficient classical simulation of such maps by considering the classical resources required to simulate these maps. Our results suggest that such maps may not be possible to simulate efficiently classically, as one would expect for non-Gaussian circuits.

\section{The cubic phase state ancilla}

\subsection{The cubic phase gate}
Gottesman, Kitaev, and Preskilll (GKP) define the (unnormalized) cubic phase state as~\cite{GKP01}
\begin{equation}
\label{eq:gamma}
| \gamma \rangle=\int \text{d} x \,\,\text{e}^{\text{i}\gamma x^3} | x \rangle
\end{equation}
and propose its employment as an off-line resource for
the Gaussian-preserving circuit depicted in Fig.~1.
Here the input state is in a position ($\hat x$) eigenstate, which is a limiting case
of displaced squeezed states, with the squeezed state given by
\begin{equation}
\label{eq:squeezedstate}
|\eta\rangle=\exp\left( \frac {\eta}{2} \hat{a}^{\dagger 2} - \frac{\eta ^*}{2} \hat{a}^2\right)|0\rangle.
\end{equation}
The state $|x\rangle$ is approximated for $\eta\in\mathbb{R}$ becoming very large and displaced
by the unitary displacement operator
\begin{equation}
\label{eq:displacement}
	D(\alpha)=\exp(\alpha\hat{a}^\dagger-\alpha^*\hat{a})
\end{equation}
with $\alpha=x\in\mathbb{R}$.
The cubic phase state, $\ket \gamma$ serves as the ancilla. The Gaussian-preserving sum-inverse
($\text{SUM}^{-1}$) gate is defined by the inverse of
$\text{SUM} = \exp(-\text{i}{\hat x}_1 {\hat p}_2)$, where the subscripts label the input and the ancilla channels respectively.
\begin{figure}
\centerline{{\epsfig{file=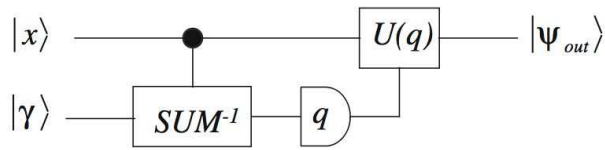,width=80mm}}}
\caption{Circuit for implementing a cubic phase gate on an input $x$ eigenstate using an ancilla cubic phase state,  and Gaussian-preserving gates. The gate $U(q)$ depends on a measurement of the ancilla in the $\hat x$-basis with outcome $q$.  }\label{n}
\end{figure}

The importance of the cubic phase state is that it can serve as an off-line resource for
effecting a cubic phase gate, namely $\exp(\text{i}\gamma \hat x^3)$, on the input $\hat x$ eigenstate using only the cubic phase state, the Gaussian-preserving sum-inverse gate, a homodyne measurement of the ancilla and feedforward. After applying the $\text{SUM}^{-1}$ gate and measuring the ancilla channel in the $x$ basis to obtain outcome $q$, 
the resultant output state is
\begin{align}
\label{eq:cgate}
 \left| {\psi_\text{out} } \right\rangle
 	=& \left\langle {q} \right|\text{SUM}^{-1} \left| x \right\rangle \left| {\gamma} \right\rangle 
			\nonumber\\
	=& \int \text{d}x_2 \, \text{e}^{\text{i}\gamma x_2^3} 
		\left\langle {q} \right|\text{SUM}^{-1} \left| x \right\rangle \left| {x_2} \right\rangle\nonumber\\	
 =& \text{e}^{\text{i} \gamma (x+q)^3}\ket{x}.
 \end{align}
Then the cubic phase gate can be effected by applying a quadratic (i.e.\ Gaussian-preserving) $q$-dependent Hamiltonian
\begin{equation}
\label{eq:Hcube}
\hat H= 3q\gamma (\hat x^2 + q \hat x + q^2/3),
\end{equation}
which generates the unitary transformation
\begin{equation}
\label{eq:Ucube}
U(q) = \exp\left\{\text{i}\left[\gamma\hat{x}^3 - \gamma\left(\hat{x}+q\right)^3\right]\right\}.
\end{equation}
This unitary operator acting on the output state (4) yields $\text{e}^{\text{i}\gamma \hat x^3} \ket{x}$.

\subsection{GKP scheme for preparation of the cubic phase state}
The cubic phase gate, plus the application of linear optics and squeezing operations, homodyne measurements and feed forward, is sufficient to generate a set of gates for universal CV quantum information processing
by optical means.  However the ideal cubic phase state defined in Eq.~(1) is not normalizable and 
therefore does not correspond to a physical state.
GKP address this issue by proposing a scheme,
depicted in Fig.~2,
to prepare an approximation to the cubic phase state 
via displacement and photon counting on one channel, or mode, of a two-mode squeezed state;
the other mode approximates a cubic phase state with a parameter conditioned by
the photon counting measurement outcome on the other state.
The two-mode squeezed state is prepared from the vacuum state by the unitary
transformation
\begin{equation}
\label{eq:2squeezedstate}
	S_{ij}(z)=\exp\left\{\frac{z}{2} \hat{a}_i^{\dag}\hat{a}_j^{\dag} -\frac{z^*}{2}\hat{a}_i\hat{a}_j\right\},
\end{equation}
analogous to the single-mode squeezed state explained in Eq.~(\ref{eq:squeezedstate}).
We analyze the scheme in detail here.

Consider the circuit shown in Fig.~2. We start with two input vacuum states, which then undergo two-mode squeezing $S_{12}(r)$ characterized by a real squeezing parameter $z=2r\in\mathbb{R}$
in Eq.~(\ref{eq:2squeezedstate}).
One of the two states is mixed with a coherent light field at a beam splitter,
which results in a momentum displacement $D_2(\alpha_2=\text{i}p_0)$
for $D$ given by Eq.~(\ref{eq:displacement}).

This channel is subject to photon counting with measurement outcome~$n_2$.
Initially we consider the ideal case of perfectly efficient photon counting.
For the case of ideal measurement, 
the state in the other channel conditioned on this measurement can be written in the $x$
representation as
\begin{equation}
\label{eq:psix}
\psi_1^{(n_2)}(x_1) = \langle x_1, n_2 | D_2(\alpha_2)S_{12}(r)\ket {0,0}.
\end{equation}
For the input state expressed in the $x$ representation, we have
\begin{equation}
\label{eq:psix2}
\psi_1^{(n_2)}(x_1) = \int \text{d} x_2 \,\exp\left(\text{i}\alpha_2 x_2\right) f(x_1,x_2)u_{n_2}(x_2),
\end{equation}
with $u_{n_2}(x_2)$ the Hermite Gaussian functions~\cite{HG}.
Here $f(x_1,x_2)$ is the $x$-representation of the two-mode squeezed state
\begin{align}
\label{eq:fx1x2}
f(x_1,x_2) =& N\exp\big\{-\tfrac{1}{2} [\cosh(2r)(x_1^2 +x_2^2)
		\nonumber	\\
	&+ 2\sinh(2r)x_1x_2 ]\big\}
\end{align}
with~$N$ a normalization constant.

GKP approximate the integral in Eq.~(9) in the large~$r$, large~$n_2$ limit as
\begin{equation}
	\psi_1^{(n_2)}(x_1) \approx \text{e}^{\text{i} \gamma x_1^3},
\end{equation}
with
\begin{equation}
\gamma = \frac{1}{6\sqrt{2n_2+1}}\;.
\end{equation}

\subsection{Exact analysis of GKP preparation scheme}
We compute an exact expression for the state $\psi_1^{(n_2)}(x_1)$  prepared via the GKP scheme (for convenience we call this the GKP state). We use the generating function for the Hermite Gaussian (see Appendix) so that
\begin{equation}
 \psi_1^{(n_2)}(x_1)= \frac{\partial ^{n_2}}{\partial t^{n_2}}   \text{e}^{-t^2}
\int \text{d}x_2 \, \text{e}^{2x_2t-x_2^2/2}
	\text{e}^{\text{i}\alpha_2 x_2} f(x_1,x_2)
\bigg |_{t=0}.
 \end{equation} 
As $f(x_1,x_2)$ is a Gaussian function of $x_2$, the above integral can now be calculated. After some manipulation, the $n_2^\text{th}$-order derivative of this integral results in an expression for the state given by
\begin{equation}
\psi_1^{(n_2)}(x) = N \text{e}^{ - \text{i}p_0 x \tanh r } \sum\limits_{m = 0}^{n_2} g_m(x),
\end{equation}
with
\begin{equation}
g_m(x) =  \left( {\begin{array}{*{20}c}
   {n_2}  \\
   m  \\
\end{array}} \right)
	(-\tanh r)^m \left( \text{i}p_0 \text{sech}^2 r \right)^{n_2  - m} u_m (x).
\end{equation}
\begin{figure}
\centerline{{\epsfig{file=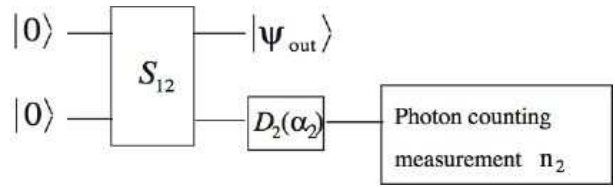,width=80mm}}}
\caption{Circuit for preparing a cubic phase state}\label{n}
\end{figure}

\subsection{Comparison between GKP state  and ideal cubic phase state}
To scrutinize the complicated state in Eq.~(14),
we first examine the $m=n_2$ term of the sum in Eq.~(14).  For large $n_2$, far from the turning points, we can use a semi-classical approximation for $u_{n_2}(x)$ (see Appendix).
Then, close to $x=0$, if $p_0 \text{sech}^2r \ll 1$ the $m=n_2$ term dominates and the state can then be approximately written as
\begin{equation}
\psi^{(n_2)}(x_1) \propto \text{e}^{\text{i}\gamma x_1^3} + \text{e}^{\text{i}\beta x_1}
\end{equation}
with $\gamma$ given by Eq.~(12) and $\beta = 2\sqrt{2n_2+1}$.  

The required cubic phase term thus arises from the $m=n_2$ term in Eq.~(14). When other terms in the sum are included, they can act to reduce the contributions of the extra linear phase term in the large $r$ large $n_2$ limit, thus leading to an approximate cubic phase state. 
We perform a numerical analysis to study the conditions for which the prepared state approaches a cubic phase state by computing the corresponding Wigner function~\cite{Wigner,Weyl} given by
\begin{equation}
W(x,p)=\int \text{d}y \, \psi^*(x-y)\psi(x+y) \text{e}^{-2\text{i}py}.
\end{equation}
The Wigner function of the ideal cubic phase state is
\begin{equation}\label{WIdeal}
	W(x,p)=2\pi N \left|\frac{4}{3\gamma}\right|^{1/3}
		\text{Ai}\left( \left[\frac{4}{3\gamma}\right]^{1/3}[3\gamma x^2 - p]\right),
\end{equation}
with $\text{Ai}(x)$ the Airy function and N a normalization factor. The Airy function causes the Wigner function to oscillate, resulting in $\theta$-dependent negative regions with $\theta = \text{tan}^{-1}(p/x)$ (Fig.~3).  

The Wigner function of the prepared state described by  Eq.~(14) in the regime of large squeezing and large $n_2$ is a sum of the Airy function of Eq.~(18) obtained from the $m=n_2$ term and additional functions arising from the additional terms in Eq.~(14). In order to examine the contribution of these terms for a finite amount of squeezing, we compute the Wigner function of the prepared state. Fig.~4 shows the effect of the squeezing parameter on the negativity of the Wigner function of the prepared state along the $x=0$ line for $n_2\approx50$. The data show that, for currently achievable levels of entanglement~\cite{S1,S2,S3,S4}, the oscillations in  the Wigner function are strongly suppressed and the state does not closely approximate a cubic phase state. 
\begin{figure}
\centerline{{\epsfig{file=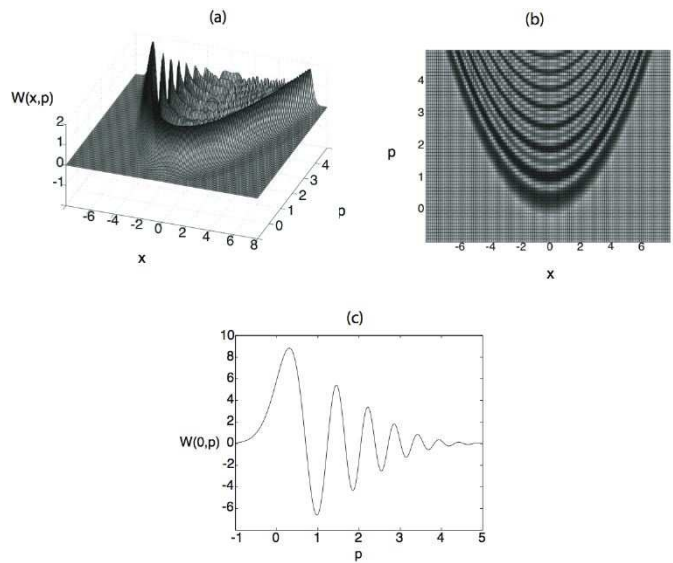,width=90mm}}}
\caption{(a) Wigner function of the ideal cubic phase state for $\gamma=0.05$. (b) Overhead view. (c) $x=0$ slice. }\label{Wp}
\end{figure}
Furthermore, the analysis thus far assumed perfect efficiency of the photodetector in channel~2 in the circuit of Fig.~2. In reality one must take into account an imperfect photodetector.

If the photodetector has an efficiency $\eta$, then the output state conditioned on a measurement of $n_2$ photons in channel~2 is a mixed state that can be written as~\cite{Bernoulli}
\begin{equation}
\rho  _1 = \sum^{\infty}_{n=n_2} P_n |\psi_1^{(n)}\rangle \langle \psi_1^{(n)}|,
\end{equation}
with
\begin{equation}
P_n=\frac{n!}{n!(n-n_2)!} \eta^{n_2}(1-\eta)^{(n-n_2)}
\end{equation}
the Bernoullli distribution.
Fig.~5 shows that as the detector efficiency decreases, the oscillations in the Wigner function are further washed out so that  the actual state diverges from the ideal cubic phase state.

\begin{figure}
\centerline{{\epsfig{file=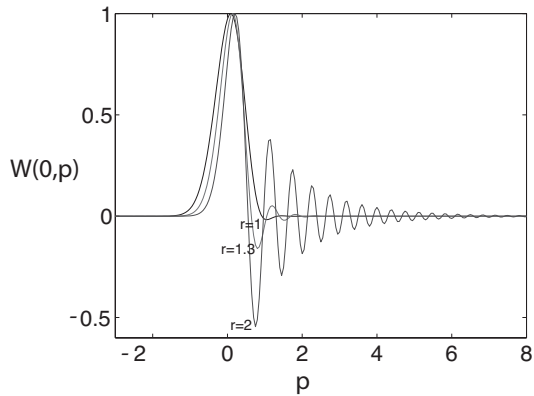, width=70mm}}}
\caption{Wigner function of the prepared state for different values of the squeezing parameter.}\label{comp}
\end{figure}

The threshold theorem of quantum error correction requires that quantum gates be implemented with a minimum fidelity~\cite{ABO97}. If we use the prepared state given by Eq.~(16) as the ancilla in the circuit of Fig.~1, we obtain a superposition of a cubic phase gate
and a quadratic transformation on the input state. In general, the non-ideal prepared ancilla state described by Eq.~(14) will result in a gate that is a superposition of the desired cubic phase gate and additional gates. The fidelity of implementing a cubic phase gate will be limited by the fidelity of preparing the cubic phase state $\ket {\gamma}$.
The ideal cubic phase state is not normalizable, making it challenging to assess the fidelity of preparing this state. Here we define the fidelity normalized over the domain~$D$ ranging from $x_{\text{min}}$ to $x_{\text{max}}$,
where the prepared state $\rho(x,x') = \langle x |\rho | x' \rangle$ has non-negligible support. For numerical evaluation, we define 
\begin{equation}
\tilde{\rho}(x,x')=\left\{
\begin{array}{cl}
\rho(x,x') &  \forall \, x, x' \in (x_{\text{min}},x_{\text{max}}) \\\
0 & \text{otherwise}
\end{array}
\right. \,.
\end{equation}
The fidelity $F$ is then calculated to be 
\begin{equation}
F=\frac{\sqrt{\langle \gamma| \tilde \rho|\gamma\rangle}}{D},
\end{equation}
with 
\begin{equation}
D=\int_{x_{\text{min}}}^{x_{\text{max}}}dx |\langle x |\gamma \rangle|^2 .
\end{equation}
We find that for a squeezing parameter $r=1.34$ and perfect detector efficiency, the normalized fidelity is roughly $20\%$.  Hence, our studies show that a close approximation to the cubic phase state cannot be prepared using current levels of squeezing and detector efficiency. 

We can define a corresponding fidelity of performing the cubic phase gate using the circuit in Fig.~2 for a given input state $|\psi\rangle$ as
\begin{equation}
F_{\text{gate}}= \sqrt{\langle \psi| U^{\dag} \sigma U |\psi\rangle}.
\end{equation}
Here $\sigma = \tilde U |\psi \rangle \langle \psi | \tilde U ^{\dag}$,
\begin{equation}
U | \psi \rangle= e^{\text{i}\gamma x^3} | \psi \rangle,
\end{equation}
and
\begin{equation}
\tilde{U} | \psi \rangle =  U(q) \left\langle {q} \right|\text{SUM}^{-1} \left| \psi \right\rangle \left| {\phi} \right\rangle
\end{equation}
with $U(q)$ given by Eq. (6) and $|\phi \rangle $ being the prepared state given by Eq.~(14). The minimum gate fidelity is the gate fidelity minimized over all input states. 
Writing the input state $|\psi \rangle$ in the $x$ basis,  the gate fidelity becomes
\begin{equation}
F_{\text{gate}}= \bigg | \int \text{d}x~|\psi(x)|^2e^{\text{-i}\gamma (x+q)^3} \psi_1^{(n_2)}(x+q) \bigg | ,
\end{equation} 
with $\psi(x) = \langle x | \psi \rangle $ and $\psi_1^{(n_2)}(x)$ given by Eq.~(14). For an input state that is an equal superposition of $x$ eigenstates, the state fidelity in Eq.~(22) is equivalent to the gate fidelity for $q$ values close to 0. Hence, as one would expect, the fidelity of preparation of the ancilla state constrains the minimum gate fidelity of the cubic phase operation.

We note that the GKP scheme for preparing the cubic phase state uses the minimal number of non-linear elements -  the key element which generates a non-linear operation is a single photodetector. Any other scheme that relied on Gaussian operations and required more photodetectors with finite efficiency may reduce the performance. However, the use of non-Gaussian ancillae in the preparation scheme may improve the preparation fidelity as discussed in the following section.

\begin{figure}
\centerline{{\epsfig{file=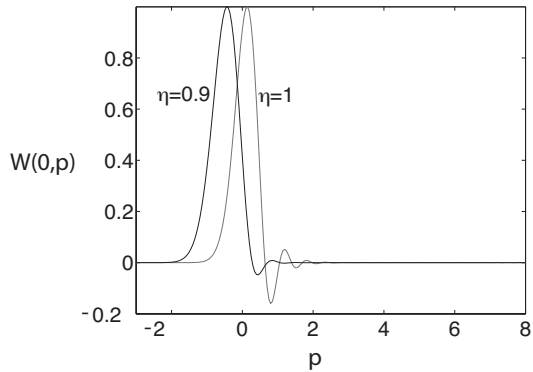, width=70mm}}}
\caption{Effect of finite detector efficiency on the Wigner function of the prepared state for $r=1.34$.}\label{comp}
\end{figure}

Whether or not the non-ideal prepared state can be
used efficiently for universal quantum information processing with Gaussian preserving maps is an
open question; perhaps a scheme using the prepared state in the circuit in Fig. 1 together with some form of error correction could provide a solution to this problem. 
A possible alternative might be to prepare many copies of the non-ideal state and perform some  purification protocol in order to prepare a state closer to the ideal cubic phase state. Furthermore, the use of alternative schemes using other non-Gaussian ancilla states such as Fock states may generate better approximations to the cubic phase state. We discuss this further in the following section on Fock states.

\section{The Fock state ancilla}
\subsection{On-line operations generated with ancilla Fock states}

\begin{figure}
\centerline{{\epsfig{file=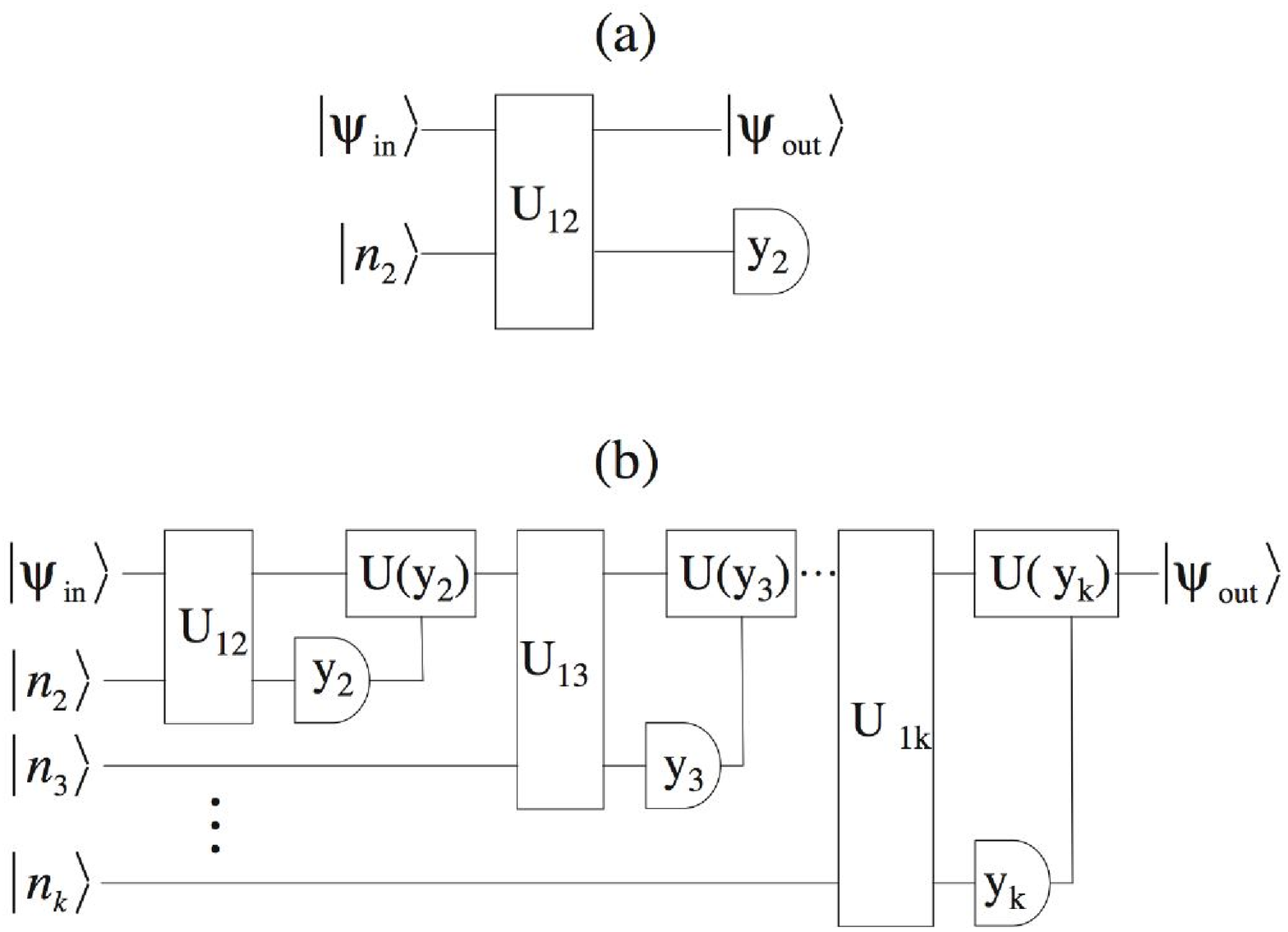,width=85mm}}}
\caption{(a)~Gaussian-preserving circuit for using a single Fock state with $n_2$ photons as an ancilla. A homodyne  measurement is performed on the ancilla with outcome $y_2$. (b)~Gaussian-preserving circuit for using $N$ photons distributed over $k-1$ channels as ancillae. The circuit is conditioned on homodyne  measurements performed on the ancillae with outcomes $y_2 \cdots y_k$.    }\label{nancilla}
\end{figure}

Given the difficulty of preparing a cubic phase state via the GKP scheme, we explore the use of other non-Gaussian states as possible ancilla states in Gaussian-preserving circuits for implementing universal CV computation. To this end, we analyze the transformations generated by input ancilla number (or Fock) states. We consider a general optical circuit as shown in Fig.~\ref{nancilla}(a). The input state $|\psi_\text{in}\rangle$ consists of $n$ electromagnetic field modes, each described as a harmonic oscillator. The field amplitude operators $\hat x$ and $\hat p$ for each mode obey the commutation relation $[\hat{x},\hat{p}]=\text{i} \hbar$. We assume that the input state is a product of Gaussian states for each of the $n$ channels. Eigenstates of $\hat x$ or $\hat p$ can be regarded as infinitely squeezed Gaussian states and are included in the set of possible input states.

We first analyze the gates generated by a single ancilla Fock state  $|n_2\rangle$  with a specific number of photons $n_2$.  The operator $U_{12}$ is any element of the Clifford group for continuous variables~\cite{BSBN02}  constructed from a combination of linear optics elements and squeezers. The output state is conditioned by a homodyne measurement of the ancilla, such that
\begin{equation}
 \left| {\psi_{\text{out}} } \right\rangle  = \left\langle {y_2} \right|U_{12} \left| {\psi_\text{in} } \right\rangle \left| {n_2} \right\rangle .
 \end{equation}
Here subscript~1 labels the input state, and subscript~2 labels the ancilla. 
Rewriting the number state $\left | n_2 \right \rangle$ as a derivative of a coherent state  $|\alpha_2 (t)\rangle$ with $\alpha_2(t)=\sqrt{2}t$ (see Appendix),
\begin{equation}
 \left| n_2 \right\rangle =  \frac{\partial ^{n_2}}{\partial t^{n_2}}  
\exp(t^2) \left| {\alpha_2(t) }  \right\rangle \bigg |_{t=0}.
\end{equation}
Inserting this into Eq.~(28), we obtain the output state
\begin{equation}
 \left| {\psi_\text{out} } \right\rangle =  \frac{\partial ^{n_2}}{\partial t^{n_2}}  
\exp(t^2) \left| {\chi(t) }  \right\rangle \bigg |_{t=0},
\end{equation}
where
\begin{equation} 
\left| {\chi(t) }  \right\rangle = \left\langle {y_2} \right|U_{12} 
	\left| {\psi_{\text{in}} } \right\rangle \left| {\alpha_2(t) } \right\rangle .
\end{equation}
The state $ \left| {\chi(t) }  \right\rangle$ is a Gaussian state since it results from a Gaussian-preserving map~\cite{BS02} with input Gaussian states undergoing CV Clifford group transformations followed by homodyne measurements. The output state in Eq.~(30) can also be derived by using the result that any linear optics circuit can be decomposed into an interferometer, followed by a parallel set of single mode squeezers, followed by another interferometer and additional displacements~\cite{B05}.

We now consider adding 
an ancilla prepared in a
product of number states  rather than a single number state. Fig.~6(b)) presents the case of  $N$ photons entering the circuit in $k-1$ channels with $\sum_{l=2}^{k} n_l = N$. Then, using the expression for a number state given by Eq.~(29), we obtain a general expression for the output state at the end of the computation,
\begin{equation}
 \left| {\psi_\text{out} } \right\rangle =\frac{\partial ^{n_{k}}}{\partial t_{k}^{n_{k}}} \cdots 
 \frac{\partial ^{n_2}}{\partial t_2^{n_2}}  
\text{e}^{t_2^2+\cdots t_{k}^2}\left| {\chi(t_2,...t_{k}) }  \right\rangle \bigg |_{t_2,...t_{k}=0},
\end{equation}
where
\begin{eqnarray} 
\left| {\chi(t_2,...t_k) }  \right\rangle &=& U_{1k} \left\langle {y_k} \right|U_{1k}...\langle y_3 |U_{13}U(y_2) \left\langle {y_2} \right|U_{12} 
	\left| {\psi_{\text{in}} } \right\rangle 
	\otimes  \left| {\alpha_2(t_2) } \right\rangle...\left| {\alpha_k(t_k) } \right\rangle 
\end{eqnarray}
is a Gaussian state resulting from Gaussian-preserving unitaries and homodyne measurements acting on the input Gaussian state. In the $x$-basis, the derivative operations in Eq.~(32) result in an output state which is a product of a Gaussian function and a polynomial function of $x$ of maximum order $2(k-1)$ for $k-1$ ancilla channels.  By changing the circuit parameters, different such functions can be generated.

\subsection{CV computation with ancilla Fock states}
It is not immediately obvious from the form of the output state in Eq.~(32), whether this circuit leads to universal CV quantum computation or not.  In order to analyze whether this circuit is sufficient for universal CV quantum computation, we only need to consider a single-mode  rather than a multi-mode input. As shown by Lloyd and Braunstein~\cite{LB99}, any non-Gaussian unitary operation on a single mode together with the set of Gaussian preserving operations is sufficient for universal CV quantum computation. This implies that since the output states in Eq~(32) are not Gaussians states, the circuit in Fig.~6 can generate a non-Gaussian operation on a single input mode and hence appears to be sufficient for CV quantum computation. The challenge then is to design a particular circuit that can generate a useful nonlinear (non-Gaussian) operation, for example a cubic phase gate on a single mode. In fact, our analysis of the GKP preparation scheme for the cubic phase state allows us to identify one possible simple circuit  to perform a non-Gaussian operation on a single mode using ancilla Fock states. Consider a single mode input state and a second ancilla mode having a single ancilla Fock state with photon number $n_2$. The ancilla state is first acted on by a momentum displacement operator with the displacement $p=\sqrt{2n_2+1}$. For large $n_2$, the resulting displaced number state can be written in the $x$-basis close to $x=0$ as a superposition of a cubic phase state and a linear phase state as given in Eq.~(16). Thus such an ancilla state when fed into the circuit of Fig.~1 will generate a superposition of a cubic phase operation and a Gaussian operation on a single mode. Although this is not an ideal cubic phase operation, as noted before, together with error correction or purification, it may be sufficient for useful CV computation. 

We identified the circuit discussed above by noting that the displaced ancilla number state is none other than the $m=n_2$ term in Eq.~(14) and that the state given in Eq.~(14) is a finite superposition of Fock states that is displaced in momentum. Thus Eq.~(14) shows that a finite superposition of displaced Fock states rather than a single displaced Fock state will closely approximate a cubic phase state for  large $n_2$ and $r \rightarrow \infty$,  and hence generate a cubic phase gate when fed into the circuit of Fig.~1. Although the GKP scheme for preparing this state is limited by finite squeezing, other schemes for preparing arbitrary finite superpositions of Fock states~\cite{D99,FGC05,PMZK93,LE96,KHR02,BDSW03} may be able to overcome the squeezing constraint. Whereas optical schemes for preparing an arbitrary, finite superposition of Fock states are non-deterministic~\cite{D99,FGC05}, the interaction of light with atoms in a cavity can deterministically generate arbitrary single mode optical fields~\cite{PMZK93,LE96,KHR02} and even Fock states with large photon numbers~\cite{BDSW03}. The feasibility of preparing a good approximation to the cubic phase state via such schemes is the topic of future work.

\subsection{Analysis of classical simulation of Gaussian circuits with ancilla Fock states}  
It is widely believed that universal quantum computation cannot be classically simulated efficiently. 
In order to further test whether the circuits in Fig.~6 can perform universal CV quantum computation,
we perform a natural adaptation of the efficient classical algorithm for simulating evolution when the circuit inputs (including ancillae) are all Gaussian states~\cite{BSBN02} to the case of Fock state ancillae. We find that the inclusion of Fock state ancillae renders this algorithm inefficient, which would be consistent with the conditions for universal quantum computation. 
As shown in Fig.~6 we look at the case of $n$ input channels initially in a product of Gaussian states along with $N$ total photons shared in a product state among $k-1$ ancilla channels.   From~\cite{BSBN02}, we know that the Gaussian-preserving map  that transforms the input state $|\psi_\text{in} \rangle$ into the state  $ \left| {\chi }  \right\rangle$ in Eq.~(28) can be efficiently simulated classically by simply following the evolution of the means and covariance matrix of the input state. 
For $n$ input channels this involves tracking a polynomial number $O(n^2)$ of real values. In the computational ($x$) basis, the state $ \left| {\chi }  \right\rangle$ is thus a Gaussian function whose coefficients are easily computed from the means and covariances. 

Assuming that the subsequent derivative operations can be performed efficiently, we seek to find the resources required to characterize the final output state after all the derivative operations have been performed. The derivative operations result in a product of the Gaussian function and a polynomial function of order $2N$ for $N$ ancilla photons. It is simple to see that the maximum number of coefficients required to specify a general polynomial of order $2N$ in $n$ variables (for $n$ input channels), is given by 
\begin{equation}
C_n =2{\sum}_{l=1}^{2N}{ \left( \begin{array}{c}
n+l-1 \\ l
\end{array}\right)}.
\end{equation}
Numerical calculations verify that this quantity scales exponentially with the number of ancilla photons and input modes. 

Our results indicate that Fock states may be a useful resource for CV quantum computation. The goal of efficiently creating Fock states on demand even at the level of single photons is thus of primary importance both for discrete as well as CV quantum information processing with linear optics. However, the design of simple gates with Fock state ancillae is not obvious so even if Fock state ancillae provided a resource for universal quantum computation, circuit design remains a challenge.

\section{Conclusions}

We have analyzed the off-line preparation and use of non-Gaussian states of light as ancillary resources in optical circuits for generating nonlinear gates for universal quantum computation. Our studies of the scheme to practically prepare a cubic phase state reveal that a good approximation to the ideal cubic phase state is not achievable with the best known current levels of squeezing and photodetectors. Nevertheless, the non-ideal prepared state can generate interesting superpositions of Gaussian and nonlinear gates that, together with error correction protocols and purification techniques, may be sufficient for universal quantum information processing. The question of designing more practical schemes for preparing and using this state remains open and requires further investigation. 

We have also performed a detailed analysis of ancilla Fock states in optical circuits, and obtained the general set of gates that can be implemented with such states. Our results indicate that a circuit using input Gaussian states, ancilla Fock states and consisting of Clifford group elements, homodyne measurements and feed forward could generate the nonlinearity required for CV computation,  but design of circuits for performing nonlinear gates is not trivial, and needs further investigation.  We also find that a classical simulation of the states generated with an ancilla  which is a product of ancilla Fock states can scale poorly with the number of ancilla photons and input modes.
This result extends previous CV quantum information no-go theorems~\cite{BSBN02,BS02}
and implies that further analysis of the transformations accessible with ancilla Fock states is of interest to assess their use for specific information processing protocols.

\section*{Acknowledgements} 
This project was supported by Alberta's Informatics Circle of Research Excellence (iCORE),
and SG was supported by an Alberta Ingenuity Fellowship and an NSERC Discovery grant.
We appreciate valuable discussions with S. Bartlett, J. Eisert, N. L\"utkenhaus and R. Stock.

\appendix*
\section{}
\section{Number states in terms of Gaussian states}

The number state $\left| n \right\rangle$ can be written in the $\hat x$-basis as
\begin{equation}
 \left| n \right\rangle  = \int \text{d}x \, u_{n}(x) 
\left| {x } \right\rangle  ,
 \end{equation} 
where $u_{n}(x)$ is a Hermite Gaussian function describing the $n^{\text{th}}$ eigenfunction of a harmonic oscillator~\cite{HG}. Using the relation
\begin{equation}
u_{n}(x)=  \text{e}^{-x^2/2}\frac{\partial ^{n}}{\partial t^{n}}   \text{e}^{-t^2+2xt}\bigg |_{t=0},
\end{equation}
and interchanging the order of the integral and the derivative,
the above equation can be rewritten as 
\begin{eqnarray}
| n \rangle &=&  \frac{\partial ^{n}}{\partial t^{n}}   \text{e}^{-t^2}
\int \text{d}x \text{e}^{2xt-x^2/2} \left| {x } \right\rangle \bigg |_{t=0}\nonumber\\
&=&\frac{\partial ^{n}}{\partial t^{n}}   \exp(t^2)
\int \text{d}x \text{e}^{-\frac{1}{2}(x-2t)^2} \left| {x } \right\rangle \bigg |_{t=0}.
 \end{eqnarray} 
Recognizing that the integral is a representation of a coherent state $|\alpha (t)\rangle$ with $\alpha(t)=\sqrt{2}t$, we can express the number state as
\begin{equation}
 \left| n \right\rangle =  \frac{\partial ^{n}}{\partial t^{n}}  
\exp(t^2) \left| {\alpha(t) }  \right\rangle \bigg |_{t=0}.
\end{equation}
For large $n$, far from the turning points, we can use a semi-classical approximation for $u_{n}(x)$ to write
\begin{equation}
u_{n}(x) =\frac{1}{\sqrt{2\pi p_n(x)}}[\text{e}^{\text{i}\int^xdyp_n(y)}+\text{e}^{-\text{i}\int^xdyp_n(y)}],
\end{equation}
with $p_n(x)=\sqrt{2n+1-x^2}$.

\end{document}